\documentclass[a4paper]{PoS}

\usepackage{amsmath}
\usepackage{braket}
\usepackage{graphicx}

\newcommand{\ee}[0]{\mathrm{e}}
\newcommand{\ii}[0]{\mathrm{i}}
\newcommand{\adjoint}[1]{\smash{\overline{#1}}\vphantom{#1}}

\DeclareMathOperator{\tr}{tr}
\DeclareMathOperator{\SU}{SU}

\title{Electromagnetic Form Factors for the $\Lambda$(1405)}

\ShortTitle{Electromagnetic Form Factors for the $\Lambda$(1405)}

\author{\speaker{Benjamin J.\ Menadue}\\
        Special Research Centre for the Subatomic Structure of Matter,\\
        School of Chemistry \& Physics, University of Adelaide, SA 5005, Australia, and \vspace{2pt}\\
        National Computational Infrastructure,\\
        Australian National University, Canberra, ACT 0200, Australia\vspace{2pt}\\
        E-mail: \email{ben.menadue@adelaide.edu.au}}

\author{Waseem Kamleh\\
        Special Research Centre for the Subatomic Structure of Matter,\\
        School of Chemistry \& Physics, University of Adelaide, SA 5005, Australia\vspace{2pt}\\
        E-mail: \email{waseem.kamleh@adelaide.edu.au}}

\author{Derek B.\ Leinweber\\
        Special Research Centre for the Subatomic Structure of Matter,\\
        School of Chemistry \& Physics, University of Adelaide, SA 5005, Australia\vspace{2pt}\\
        E-mail: \email{derek.leinweber@adelaide.edu.au}}

\author{M.\ Selim Mahbub\\
        Special Research Centre for the Subatomic Structure of Matter,\\
        School of Chemistry \& Physics, University of Adelaide, SA 5005, Australia, and\vspace{2pt}\\
        CSIRO Computational Informatics, College Road, Sandy Bay, TAS 7005, Australia\vspace{2pt}\\
        E-mail: \email{md.mahbub@adelaide.edu.au}}

\author{Benjamin J.\ Owen\\
        Special Research Centre for the Subatomic Structure of Matter,\\
        School of Chemistry \& Physics, University of Adelaide, SA 5005, Australia\vspace{2pt}\\
        E-mail: \email{benjamin.owen@adelaide.edu.au}}

\abstract{ 
  Building on our successful technique to isolate the
  otherwise-elusive $\Lambda$(1405) using correlation matrix techniques and
  multiple source and sink smearings, we present calculations of the
  quark sector contributions to the electric form factors of the
  $\Lambda$(1405). Using the PACS-CS $(2+1)$-flavour full-QCD ensembles
  available through the ILDG, our calculations reveal behaviour
  consistent with the development of a non-trivial molecular
  $\overline{K}N$ bound-state component as one approaches the physical
  values of the $u$ and $d$ quark masses.
}

\FullConference{31st International Symposium on Lattice Field Theory - LATTICE 2013\\
		July 29 - August 3, 2013\\
		Mainz, Germany}

\begin{document}

\section{Introduction}

\subsection{Motivation}

The $\Lambda$(1405) is the lowest-lying odd-parity state of the $\Lambda$ baryon.
With a mass of $1405.1^{+1.3}_{-1.0}$ MeV, it is lower than the lowest
odd-parity state of the non-strange nucleon (N(1535)). We now
understand this unusually low mass to be a consequence of its
flavour-singlet structure
\cite{Menadue:2012kc,Menadue:2011zz,Menadue:2011zza}.

While early lattice QCD studies of this state were unable to reproduce
the associated mass suppression, an extrapolation of the trend of the
lowest-lying state in our recent work \cite{Menadue:2011pd} was
consistent with the physical mass of the $\Lambda$(1405).  The use of a
correlation matrix is key to this success.  Away from the
$\SU(3)$-flavour--symmetry limit, octet and singlet state components are
strongly mixed.  A correlation matrix analysis is required to isolate
the QCD eigenstates and reveal the non-trivial mixing of octet and
singlet components.  Subsequent studies have confirmed our results
\cite{Engel:2012qp}. 

\begin{figure}[b]
\begin{center}
\includegraphics[width=0.49\textwidth]{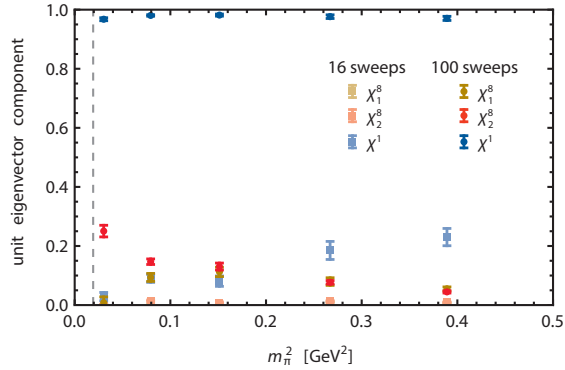}
\end{center}
\vspace{-15pt}
\caption{\label{fig:ev}Relative strength of each interpolating field's
  eigenvector component, $u_i(\mathbf{0})$ of Eq.~(2.4), for the
  $\Lambda$(1405) as a function of $m_\pi^2$.  The light-coloured points
  are smeared with 16 sweeps of smearing and the dark points with 100
  sweeps. The gold points correspond to the flavour-octet operator with
  a $(q\,C\,\gamma_5\,q)\,q$ Dirac structure, the red points correspond
  to the same flavour-octet structure but with a
  $(q\,C\,q)\,\gamma_5\,q$ Dirac structure.  The blue points
  correspond to the flavour-singlet operator.
}
\end{figure}

Key to our work's success is the inclusion of both multiple source and
sink smearings, and multiple flavour and Dirac structures in the
choice of interpolating fields.  Figure~\ref{fig:ev} displays the
flavour composition of the $\Lambda$(1405) in the relative components of the
different interpolating fields from a $6 \times 6$ correlation matrix
as the pion-mass varies.  While the highly-smeared, flavour-singlet
operator is always the dominant contribution, an octet component
becomes important away from the $\SU(3)$-flavour--symmetry limit.  It's
also interesting to observe the smaller role for the 16-sweep-smeared
flavour-singlet interpolator as the $u$ and $d$ quark masses become light.

\begin{figure}
\begin{center}
\includegraphics[width=0.49\textwidth]{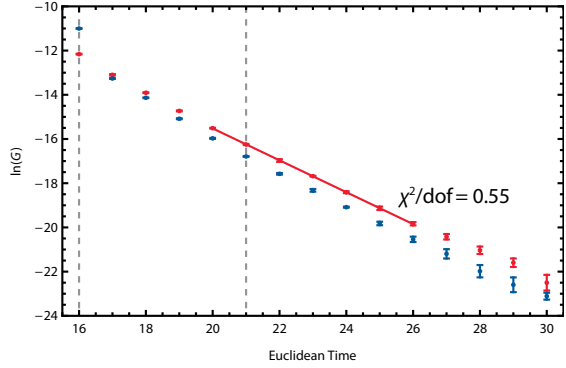}
\end{center}
\vspace{-15pt}
\caption{\label{fig:cmcf}Comparison between the correlation functions
  extracted from a variational analysis (red) and from a traditional
  analysis (blue). The variational analysis results in a lower slope
  that is constant over a larger fit window, indicating a more complete
  isolation of the lowest-lying state. The fermion source is at $t = 16$
  (first dashed line), and the current insertion for the form factor
  analysis is at $t = 21$ (second dashed line).}
\end{figure}

The variational analysis is also necessary to obtain a signal that is
stable for sufficiently long Euclidean times after the fermion source
(in our case, $t = 16$) to perform the current insertion for the form
factor analysis. Figure~\ref{fig:cmcf} demonstrates the long-term
stability of the correlation function extracted from the variational
analysis. We insert the current at $t = 21$, five time slices after
the fermion source.  Here the variational analysis provides optimal
coupling to the lowest state and effective suppression of excited
state contaminations.  Moreover, the two-point correlation function
remains stable long enough after the current insertion to extract
reasonable measures of the form factors.

\subsection{Simulation Details}

We use the PACS-CS $(2+1)$-flavour full-QCD ensembles
\cite{Aoki:2008sm}, made available through the ILDG
\cite{Beckett:2009cb}. These ensembles have a lattice size of $32^3
\times 64$ with $\beta = 1.90$, and there are five pion masses ranging
from 640\,MeV down to 156\,MeV. Each ensemble has the same dynamical
strange quark mass, corresponding to $\kappa_\textrm{s} = 0.13640$,
however we perform our simulations using $\kappa_\textrm{s} = 0.13665$
for the valence strange quarks to reproduce the correct kaon mass in
the physical limit. As demonstrated in Ref.~\cite{Menadue:2012kc}, this
partial quenching is subtle and doesn't affect the extracted form
factors. We consider setting the scale using both the Sommer and
PACS-CS schemes, however we find no qualitative difference between them
in the behaviour of the form factors.  We select the PACS-CS scheme in
the following. 

\section{Techniques}

\subsection{Two-Point Variational Analysis}

To extract correlation functions from a variational analysis
\cite{Leinweber:2004it}, we first need to construct the correlation
matrix. If we consider some set of operators $\set{\chi_i}$ that
couple to the states of interest, the associated correlation matrix
can be written as
\begin{equation}
G_{ij}(\Gamma; \mathbf{p}; t) = \sum_{\mathbf{x}} \ee^{-\ii \, \mathbf{p} \cdot \mathbf{x}} \tr (\Gamma \braket{\Omega | \chi_i(x) \adjoint{\chi}_j(0) | \Omega}),
\end{equation}
where $\Gamma$ is some Dirac matrix that sensibly selects the
appropriate components of the resultant spinor matrix. We then solve
for the left ($\mathbf{v}^\alpha(\mathbf{p})$) and right
($\mathbf{u}^\alpha(\mathbf{p})$) generalised eigenvectors of
$G(\Gamma; \mathbf{p}; t + \delta t)$ and $G(\Gamma; \mathbf{p}; t)$,
so that 
\begin{align}
G(\Gamma; \mathbf{p}; t + \delta t) \, \mathbf{u}^\alpha(\mathbf{p}) &= \ee^{-E_\alpha(\mathbf{p})\,\delta t} \, G(\Gamma; \mathbf{p}; t) \, \mathbf{u}^\alpha(\mathbf{p}),\textrm{ and} \\
\mathbf{v}^{\alpha\top}(\mathbf{p}) \, G(\Gamma; \mathbf{p}; t + \delta t) &= \ee^{-E_\alpha(\mathbf{p})\,\delta t} \, \mathbf{v}^{\alpha\top}(\mathbf{p}) \, G(\Gamma; \mathbf{p}; t).
\end{align}

These eigenvectors identify the ``ideal'' combinations $\phi^\alpha$
of the original operators $\chi_i$ that perfectly isolate individual
energy eigenstates at momentum $\mathbf{p}$. As such, we can write 
\begin{equation}
\phi^\alpha(\mathbf{p}) = v_i^\alpha(\mathbf{p}) \, \chi_i \qquad
\adjoint{\phi}^\alpha(\mathbf{p}) = \adjoint{\chi}_i \,
u_i^\alpha(\mathbf{p}) \, .
\label{phi_definition}
\end{equation}
Note that the Greek indices, $\alpha$ and $\beta$, label states and
are not to be summed over; unlike the Latin operator indices ($i,
j, \ldots$) which are summed over.  Using these operators, we can
extract correlation functions for the individual eigenstates,
\begin{align}
G_\alpha(\Gamma; \mathbf{p}; t) &= \sum_{\mathbf{x}} \ee^{-\ii \, \mathbf{p} \cdot \mathbf{x}} \tr (\Gamma \braket{\Omega | \phi^\alpha(x) \adjoint{\phi}^\alpha(0) | \Omega}) \\
&= \sum_{\mathbf{x}} \ee^{-\ii \, \mathbf{p} \cdot \mathbf{x}} \tr (\Gamma \braket{\Omega | v_i^\alpha(\mathbf{p}) \, \chi_i \, \adjoint{\chi}_j \, u_j^\alpha(\mathbf{p}) | \Omega}) \\
&= v_i^\alpha(\mathbf{p}) \left( \sum_{\mathbf{x}} \ee^{-\ii \, \mathbf{p} \cdot \mathbf{x}} \tr (\Gamma \braket{\Omega | \chi_i \, \adjoint{\chi}_j | \Omega}) \right) u_j^\alpha(\mathbf{p}) \\
&= \mathbf{v}^{\alpha\top}(\mathbf{p}) \, G(\Gamma; \mathbf{p}; t) \, \mathbf{u}^\alpha(\mathbf{p}).
\end{align}

\subsection{Three-point Variational Analysis}

To extract the form factors for an energy eigenstate $\alpha$, we need
to calculate the three-point correlation function 
\begin{equation}
G_\alpha^\mu(\Gamma; \mathbf{p}', \mathbf{p}; t_2, t_1) = \sum_{\mathbf{x}_1, \mathbf{x}_2} \ee^{-\ii \, \mathbf{p}' \cdot \mathbf{x}_2} \, \ee^{\ii (\mathbf{p}' - \mathbf{p}) \cdot \mathbf{x}_1} \tr ( \Gamma \braket{ \Omega | \phi^\alpha(x_2) \, j^\mu(x_1) \, \adjoint{\phi}^\alpha (0) | \Omega} ).
\end{equation}
where $j^\mu$ is the current. This takes the form
\begin{equation}
G_\alpha^\mu(\Gamma; \mathbf{p}', \mathbf{p}; t_2, t_1) = \ee^{-E_\alpha(\mathbf{p}')\,(t_2 - t_1)} \, \ee^{-E_\alpha(\mathbf{p}) t_1} \tr \left(\Gamma \, \sum_{s,s'} \braket{\Omega | \phi^\alpha | p', s'} \braket{p', s' | j^\mu | p,s} \braket{p, s | \adjoint{\phi}^\alpha | \Omega}\right)
\end{equation}
where the current matrix element $\braket{p', s' | j^\mu | p,s}$
encodes the form factors of the interaction. 

Using the nature of the ``perfect'' operators $\phi^\alpha$, we can
rewrite this perfect three-point correlation function in terms of the
non-projected three-point correlation functions $G ^\mu_{ij}$
calculated using the original operators $\chi_i$ using 
\begin{align}
G_\alpha^\mu(\Gamma; \mathbf{p}', \mathbf{p}; t_2, t_1) &= \sum_{\mathbf{x}_1, \mathbf{x}_2} \ee^{-\ii \, \mathbf{p}' \cdot \mathbf{x}_2} \, \ee^{\ii (\mathbf{p}' - \mathbf{p}) \cdot \mathbf{x}_1} \tr ( \Gamma \braket{ \Omega | v_i^\alpha(\mathbf{p}') \, \chi^i(x_2) \, j^\mu(x_1) \, \adjoint{\chi}_j(0) \, u_j^\alpha(\mathbf{p}) | \Omega} ) \\
&= \mathbf{v}^{\alpha\top}(\mathbf{p}') \, G^\mu(\Gamma; \mathbf{p}', \mathbf{p}; t_2, t_1) \, \mathbf{u}^\alpha(\mathbf{p}).
\end{align}

To eliminate the temporal dependence of the three-point correlation
function, we construct the ratio \cite{Leinweber:1990dv,Boinepalli:2006xd}
\begin{equation}
R_\alpha^\mu(\Gamma', \Gamma; \mathbf{p}', \mathbf{p}; t_2, t_1) = \left( \frac{G^\mu_\alpha(\Gamma; \mathbf{p}', \mathbf{p}; t_2, t_1)\, G^\mu_\alpha(\Gamma; \mathbf{p}, \mathbf{p}'; t_2, t_1)}{G_\alpha(\Gamma'; \mathbf{p}'; t_2) \, G_\alpha(\Gamma'; \mathbf{p}; t_2)} \right)^{1/2},
\end{equation}
and then to further simplify things define a reduced ratio as
\begin{equation}
\adjoint{R}^\mu_\alpha(\Gamma', \Gamma; \mathbf{p}', \mathbf{p}; t_2, t_1) = \left( \frac{2E_\alpha(\mathbf{p})}{E_\alpha(\mathbf{p}) + m_\alpha} \right)^{1/2} \left( \frac{2E_\alpha(\mathbf{p}')}{E_\alpha(\mathbf{p}') + m_\alpha} \right)^{1/2} R_\alpha^\mu(\Gamma', \Gamma; \mathbf{p}', \mathbf{p}; t_2, t_1)
\end{equation}

\subsection{Choice of Operators}

Since the $\Lambda$ baryon lies in the centre of the $\SU(3)$-flavour symmetry
group, there are many operators that will couple to it.  We can
consider interpolating fields having either a flavour-octet or
-singlet symmetry structure, in addition to the usual two Dirac
structures.  (We note that the two Dirac structures are related
through a Fierz transformation for the flavour-singlet
operator.)  These operators have the forms \cite{Leinweber:1990dv}
\begin{align}
\chi_1^8 &= \frac{1}{\sqrt{6}} \varepsilon^{abc} \left( 2(u^a \, C \, \gamma_5 \, d^b) \, s^c + (u^a \, C \, \gamma_5 \, s^b) \, d^c - (d^a \, C \, \gamma_5 \, s^b) \, u^c \right), \\
\chi_2^8 &= \frac{1}{\sqrt{6}} \varepsilon^{abc} \left( 2(u^a \, C \, d^b) \, \gamma_5 \, s^c + (u^a \, C \, s^b) \, \gamma_5 \, d^c - (d^a \, C \, s^b) \, \gamma_5 \, u^c \right),\textrm{ and} \\
\chi^1 &= 2 \varepsilon^{abc} \left( (u^a \, C \, \gamma_5 \, d^b) \, s^c - (u^a \, C \, \gamma_5 \, s^b) \, d^c + (d^a \, C \, \gamma_5 \, s^b) \, u^c \right),
\end{align}
where we have suppressed the $x$ dependence for clarity. We also
expand our operator basis by including operators smeared by differing
amounts of gauge-invariant Gaussian smearing \cite{Gusken:1989qx}. 

Note that if too few operators are included, the states won't be
sufficiently isolated, while if we include too many the correlation
matrix will be too ill-conditioned to solve for the generalised
eigenvectors. We focus on the $6 \times 6$ matrix formed by using
$\chi_1^8$, $\chi_2^8$, and $\chi^1$ together with 16 and 100 sweeps
of smearing. The selection of these two smearings gives results
consistent with other smearing combinations involving at least one of
100 or 200 smearing sweeps, but offer reduced statistical noise.

\subsection{Extracting Form Factors}

Once we have calculated the reduced ratio $\adjoint{R}^\mu_\alpha$ for
some energy eigenstate $\alpha$, we can turn our attention to
extracting information such as the form factors of the
interaction. The current matrix element for spin-$1/2$ baryons can be
written in the form
\begin{equation}
\braket{p', s' | j^\mu | p, s} =
\left(\frac{m_\alpha^2}{E_\alpha(\mathbf{p}) E_\alpha(\mathbf{p}')}
\right)^{1/2} \adjoint{u} \left( F_1(q^2) \, \gamma^\mu + \ii \,
F_2(q^2) \, \sigma^{\mu\nu} \frac{q^\nu}{2m_\alpha} \right) u \, ,
\end{equation}
where $F_1$ and $F_2$ are the Dirac and Pauli form factors. These are
related to the Sachs form factors through 
\begin{align}
\mathcal{G}_\textrm{E}(q^2) &= F_1(q^2) - \frac{q^2}{(2m_\alpha)^2}
F_2(q^2) \, , \textrm{ and} \\
\mathcal{G}_\textrm{M}(q^2) &= F_1(q^2) + F_2(q^2) \, .
\end{align}

A suitable choice of momentum $\mathbf{q}$ and the Dirac matrices
$\Gamma$ and $\Gamma'$ allows us to directly access the Sachs form
factors through an ``effective'' form factor
\cite{Leinweber:1990dv}. In particular, we have 
\begin{align}
\mathcal{G}_\textrm{E}^{\textrm{eff},\alpha}(q^2) &=
\adjoint{R}^\mu_\alpha(\Gamma_4^\pm, \Gamma_4^\pm; \mathbf{q},
\mathbf{0}; t_2, t_1)\, , \textrm{ and} \\
|\varepsilon_{ijk} \,
q^i|\mathcal{G}_\textrm{M}^{\textrm{eff},\alpha}(q^2) &=
(E_\alpha(\mathbf{q}) + m_\alpha)\adjoint{R}^\mu_\alpha(\Gamma_j^\pm,
\Gamma_4^\pm; \mathbf{q}, \mathbf{0}; t_2, t_1)\, ,
\end{align}
where the appropriate Dirac matrices are
\begin{equation}
\Gamma_j^+ = \frac{1}{2}\begin{bmatrix}\sigma_j & 0 \\ 0 &
  0\end{bmatrix}\, , \qquad 
\Gamma_4^+ = \frac{1}{2}\begin{bmatrix}\mathbb{I} & 0 \\ 0 &
  0\end{bmatrix} \, ,
\end{equation}
for the positive-parity states and
\begin{equation}
\Gamma_j^- = -\gamma_5 \Gamma_j^+ \gamma_5 =
\frac{1}{2}\begin{bmatrix}0 & 0 \\ 0 & \sigma_j\end{bmatrix}\, ,
  \qquad 
\Gamma_4^- = -\gamma_5 \Gamma_4^+ \gamma_5 =
\frac{1}{2}\begin{bmatrix}0 & 0 \\ 0 & \mathbb{I}\end{bmatrix} \, ,
\end{equation}
for negative parity.

Since the momentum transfer for a constant current momentum depends on
the mass of the target, to obtain accurate comparisons we need to
correct for any subtle changes in $Q^2$. To do this, we assume that
$\mathcal{G}_\textrm{E}$ has a dipole dependence on $Q^2$, so that 
\begin{equation}
\mathcal{G}_\textrm{E}(Q^2) = \left( \frac{\Lambda}{\Lambda + Q^2}
\right)^2 \mathcal{G}_\textrm{E}(0) \, .
\end{equation}
We can solve for $\Lambda$ by taking our measured
$\mathcal{G}_\textrm{E}(Q^2)$ in combination with the
$\mathcal{G}_\textrm{E}(0) = 1$ implication of the unit-charge
normalisation.  This can then be used to evaluate
$\mathcal{G}_\textrm{E}$ at any nearby $Q^2$. Since there is little
variation in $Q^2$ across the range of $m_\pi^2$ under consideration,
this is a small correction.

\section{Results}

\begin{figure}
\begin{center}
\includegraphics[width=0.48\textwidth]{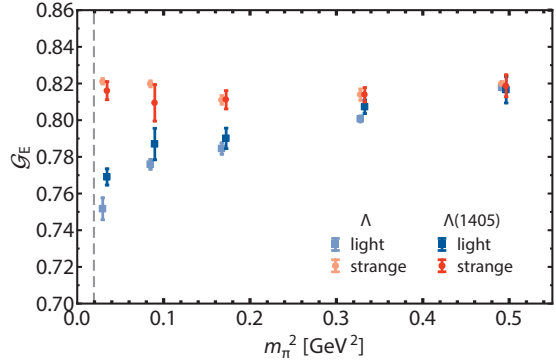}
\end{center}
\vspace{-15pt}
\caption{\label{fig:FF}Sachs electric form factors at
  $Q^2 = 0.16\ \textrm{GeV}^2/c^2$.  Results for the
  individual unit-charged quark flavour sectors for the $\Lambda$(1405) (dark
  points) are compared with those for the ground state $\Lambda$ (light).}
\end{figure}

Figure~\ref{fig:FF} presents the pion mass dependence of the Sachs
electric form factors for the individual quark sectors for both the $\Lambda$(1405)
and the ground-state even-parity $\Lambda$ at $Q^2 = 0.16$ GeV${}^2/c^2$.
We see little change between the ground state $\Lambda$ and the $\Lambda$(1405). At heavy
quark masses approaching the flavour-symmetry limit, the light ($u$ or $d$)
quarks in the $\Lambda$(1405) have the same distribution as the strange quark as
required by the singlet symmetry.  As the $u$ and $d$ quarks become light,
we observe a significant departure from the flavour symmetry, reminiscent of
Fig.~\ref{fig:ev} where the octet interpolator becomes important for
the excitation of the $\Lambda$(1405) in the light-quark region.  It is also
interesting to note that the strange quark form factor variation is a
pure environment effect as the mass of the strange quark is held
fixed.

The deviation from this flavour-singlet picture as the pion mass approaches
its physical value is consistent with the development of
a $\overline{K}N$ component in the structure of the $\Lambda$(1405). If we
consider such a dressing, the centre of mass lies nearer the heavier
nucleon, so the anti-light--quark contribution is distributed further
out by the $\overline{K}$; this leaves an enhanced light-quark form
factor. Similarly, the strange quark is also distributed further out
by the $\overline{K}$ and thus results in a suppressed form factor
relative to the ground state $\Lambda$.

\section{Conclusion}
\vspace{-6pt}
Variational techniques provide a robust technique for accessing and
isolating the eigenstates of QCD on a finite volume lattice.  In the
case of the $\Lambda$(1405) the correlation matrix is vital in separating the
nearby octet and singlet states of the spectrum.  Herein, we have
presented the very first calculation of the electric form
factors of the unusual $\Lambda$(1405).  Our results are consistent with the
development of a non-trivial $\overline{K}N$ bound-state component as
one approaches the physical values of the $u$ and $d$ quark masses.

\section*{Acknowledgements}
\vspace{-6pt}
This research was undertaken with the assistance of resources at the
NCI National Facility in Canberra, Australia, and the iVEC facilities
at Murdoch University (iVEC@Murdoch) and the University of Western
Australia (iVEC@UWA). These resources were provided through the
National Computational Merit Allocation Scheme, supported by the
Australian Government, and the University of Adelaide Partner Share.
This research is supported by the Australian Research Council.

\end{document}